\documentstyle[twoside,fleqn,espcrc2,psfig]{article}

\title{Charge stripes and spin correlations in copper-oxide superconductors}

\author{J. M. Tranquada\address{Physics Department, Brookhaven National
Laboratory, Upton, NY 11973, USA}%
\thanks{Work at Brookhaven is supported by Contract No.~DE-AC02-76CH00016,
Division of Materials Sciences, U.S. Department of Energy.}}

\begin{document}

\begin{abstract}
Recent neutron diffraction studies have yielded evidence that, in a particular
cuprate family, holes doped into the CuO$_2$ planes segregate into stripes that
separate antiferromagnetic domains.  Here it is shown that such a picture
provides a quantitatively consistent interpretation of the spin fluctuations
measured by neutron diffraction in La$_{1.85}$Sr$_{0.15}$CuO$_4$ and
YBa$_2$Cu$_3$O$_{6+x}$.
\end{abstract}

\maketitle

\section{INTRODUCTION}

To obtain superconductivity in a layered copper-oxide compound, it is necessary
to introduce charge carriers into the antiferromagnetic CuO$_2$ planes.  Recent
neutron diffraction studies of the system La$_{1.6-x}$Nd$_{0.4}$Sr$_x$CuO$_4$ 
\cite{tran95a,tran97a} provide
evidence that the dopant-induced holes choose to segregate into
periodically-spaced stripes which separate antiferromagnetic domains, in a
manner similar to that found in hole-doped La$_2$NiO$_4$ \cite{chen93}.
The charge and spin stripe modulations are identified by the appearance of
scattering at incommensurate positions.  In the Nd-doped system, elastic
scattering is observed, corresponding to static stripes.  In pure
La$_{2-x}$Sr$_x$CuO$_4$, the magnetic scattering that is observed is purely
inelastic \cite{cheo91}.  Where samples with and without Nd, but with the same Sr
concentration, have been measured, the incommensurate (IC) splittings of the
magnetic signal are found to be essentially identical
\cite{tran97a,cheo91,yama96}.  It has been proposed that the spin 
correlations in the two systems are fundamentally the same, thus implying
similar charge correlations.  The static nature of the stripes in the Nd-doped
system is attributed
\cite{tran95a} to pinning of the otherwise dynamic correlations by a special
distortion of the lattice
\cite{craw91}.  That distortion is driven by purely ionic interactions and
is stabilized by the smaller ionic radius of the substituted Nd.

To strengthen the case for the charge-stripe interpretation, it is necessary
go beyond a simple comparison of incommensurabilities.  In this paper I will 
show that the neutron scattering measurements of spin fluctuations in
superconducting La$_{1.85}$Sr$_{0.15}$CuO$_4$ and YBa$_2$Cu$_3$O$_{6.6}$ are
quantitatively consistent with what one should expect based on the stripe
picture.  The general expectations are described in section 2.  Specific
analyses for the two different superconductors are given in sections 3 and 4,
respectively.  Section 5 contains some further discussion.

\section{MODEL}

First recall the situation for the 2D Heisenberg model
\cite{auer88}.  At
$T=0$, where long-range order exists, the spin-wave dispersion is linear at small
${\bf q}$ ($={\bf Q}-{\bf Q}_{\rm AF}$), and can be expressed as $\omega=c_0q$. 
At finite temperatures, thermal fluctuations result in a finite spin-spin
correlation length $\xi$.  Spin waves are overdamped for $q<1/\xi$, but spin-wave
dispersion should still be detectable for $q>1/\xi$ (as long as $kT$ is less
than the superexchange energy $J$).  The characteristic energy $\Gamma_0$
separating the over- and underdamped regimes is then given by 
\begin{equation}
  \Gamma_0\approx\hbar c_0/\xi.
\end{equation}
As for the dynamical susceptibility, $\chi''({\bf Q},\omega)$, $\Gamma_0$ acts
somewhat like a gap energy \cite{auer88}.  If $\chi''$ is integrated over 
${\bf Q}$ to give
$\tilde{\chi}''(\omega)$, then there is a peak at $\hbar\omega\approx\Gamma_0$,
with $\tilde{\chi}''(\omega)$ going to zero as $\omega$ goes to zero. 

Next consider an ordered stripe phase.  The charge stripes need not affect the
superexchange energy within the antiferromagnetic domains, but they certainly
will weaken the magnetic coupling between domains.  Castro Neto and Hone
\cite{neto96} have considered the effect of the anisotropic coupling on the spin
wave velocity.  For a large anisotropy, the velocity $c$ for propagation
parallel to the stripes is equal to $c_0/\sqrt{2}$; it is more difficult to
estimate an effective velocity perpendicular to the stripes because the stripe
modulation should result in several optical branches as well as the acoustic
mode.  An experimental test of this situation is given by recent measurements
of spin waves in stripe-ordered La$_2$NiO$_{4.133}$ \cite{tran97c}.  That study
found $c=0.6c_0$ for propagation parallel to the stripes.  Dispersion was less
well defined but comparable in the perpendicular direction.  (Note that the
spin waves disperse out of the IC points in reciprocal space that
characterize the magnetic order.)  Thus, it seems reasonable to take
$c=c_0/\sqrt{2}$ as an estimate of the effective spin-wave velocity in an
ordered stripe phase.

Finally, consider a stripe phase with a finite spin-spin correlation length
induced by fluctuations of the charge stripes.  The characteristic energy
separating over- and underdamped modes should now be
\begin{equation}
  \Gamma\approx\hbar c_0/\sqrt{2}\xi.
\end{equation}
Excitations at $\hbar\omega>\Gamma$ should look spin-wave-like.  On the other
hand, a scan at a fixed excitation energy smaller than $\Gamma$ through the
IC points characterizing the stripe correlations should show a broad
peak centered at each IC point, with the width varying inversely as
$\xi$.  To obtain an analytic formula describing the $q$-dependence of the
scattering, consider a model in which the spins have a uniform correlation
within each domain, but where the correlations between domains fall off
exponentially with distance.  If the charge stripes are separated by $n$
lattice spacings, then the intensity along ${\bf Q}=(h,\frac12,0)$ should be
proportional to
\begin{equation}
  I(h)\sim |F|^2{1-p^2\over 1+p^2-2p\cos(2\pi nh)},
\end{equation}
where $p=(-1)^{n+1}\exp(-na/\xi)$, and $F$ is the structure factor for a single
domain.  (The factor of $-1$ included in $p$ takes care of the antiphase
relationship between neighboring domains.)  For the calculations below, it will
be assumed that the spin correlations also fall off exponentially parallel to
the stripes, with the same correlation length $\xi$.

\section{La$_{1.85}$Sr$_{0.15}$CuO$_4$}

Recently, Hayden {\it et al.}\ \cite{hayd96a} reported inelastic neutron
scattering measurements of $\chi''({\bf Q},\omega)$ in
La$_{1.86}$Sr$_{0.14}$CuO$_4$, covering a very large energy range.  They
found that $\chi''$ at $\hbar\omega>100$~meV could be modelled reasonably well
using spin wave formulas, with only a slight reduction in $J$.  At lower
energies, they showed that $\tilde{\chi}''(\omega)$ has a peak at
$\Gamma=(22\pm5)$~meV.  If this latter result is combined with the value
$\hbar c_0=(850\pm30)$~meV-\AA\ \cite{aepp89}, then by inverting Eq.~(2) one
obtains the estimate $\xi=(27\pm6)\,\,{\rm \AA}=(7.0\pm1.5)a$.

\begin{figure}[htb]
\null\vspace{9pt}
\psfig{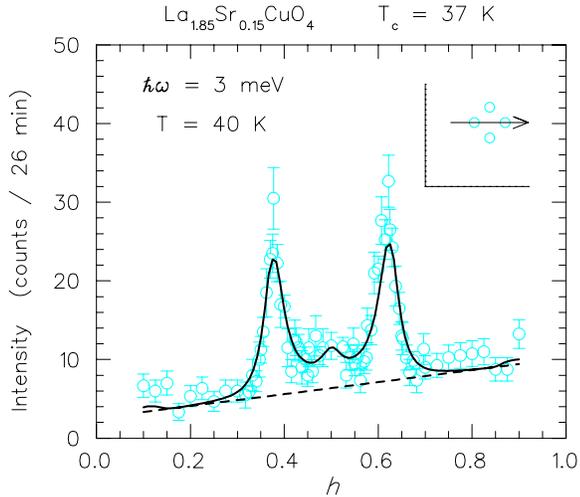}
\caption{Inelastic scan at $\hbar\omega=3$~meV through the incommensurate
points (see inset) measured at $T=40$~K on a crystal of
La$_{1.85}$Sr$_{0.15}$CuO$_4$ with $T_c=37$~K.  Solid line is the calculation
described in the text; dashed line is the fitted background.  Scan is along
${\bf Q}=(h,\frac12,0)$, within the $(hk0)$ zone, as indicated schematically in
the inset.}
\label{fg:lsco} \end{figure}

Figure~\ref{fg:lsco} shows an inelastic scan, at a fixed energy transfer of
3~meV, through the IC peaks found in a single crystal of
La$_{1.85}$Sr$_{0.15}$CuO$_4$ \cite{sternlieb}.  The peak positions are
consistent with a charge stripe period of $n=4$.  The solid line through the data
points is a calculation  using Eq.~(3) and the estimate $\xi=7a$---only the
amplitude and the linear background have been adjusted to fit the data.  (The
calculated intensity, taking into account signal from all 4 IC
peaks, was also convolved with the experimental resolution function.)  The
calculation clearly gives an excellent description of the data.

\section{YBa$_2$Cu$_3$O$_{6.6}$}

In contrast to the 214 system, measurements of $\chi''$ in
YBa$_2$Cu$_3$O$_{6.6}$ do not show well resolved IC peaks
\cite{tran92}.  As shown in Fig.~2, the scattered intensity is peaked at 
${\bf Q}_{\rm AF}$, but with a peak shape that has a rather flat top and steep
sides.  Despite the lack of clear evidence for incommensurability, it is
possible nevertheless to describe the data with the same stripe model discussed
above.  To give the stripe picture an honest test, $\xi$ will again be
estimated using Eq.~(2).  The value of $\Gamma$ is approximately $(30\pm5)$~meV
\cite{tran92}, and $\hbar c_0=(670\pm30)$~meV-\AA\ \cite{sham93,hayd96b}, which
together yield
$\xi=(16\pm3)\,\,{\rm\AA}=(4.1\pm0.8)a$.  To estimate the stripe spacing,
assuming that the hole density within the charge stripes is the same as in the
214 system \cite{tran95a}, it is first necessary evaluate the hole concentration
within a plane.  Inverting the formula \cite{pres91}
\begin{equation}
  T_c/T_{c,max} = 1-82.6(p-0.16)^2,
\end{equation}
yields $p\approx0.088$ for the $T_c=53$-K sample, which corresponds closely to
$n=6$.  The solid line through the data in Fig.~2 represents a calculation
using the same stripe model as above, with the parameter values $n=6$ and
$\xi=4.1a$.  The calculated function was convolved with the spectrometer
resolution function, and only the amplitude and background were adjusted to fit
the data.

\begin{figure}[htb]
\null\vspace{9pt}
\psfig{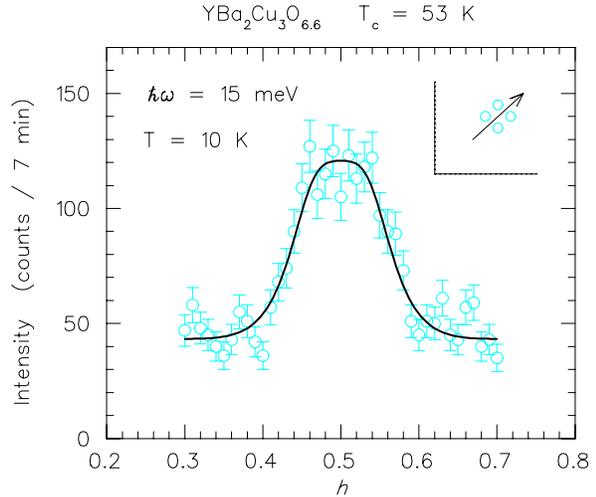}
\caption{Inelastic scan at $\hbar\omega=15$~meV along the zone diagonal (see
inset) measured at $T=10$~K on a crystal of YBa$_2$Cu$_3$O$_{6.6}$ with
$T_c=53$~K.  The solid line is the calculation described in the text.  The scan
was measured along ${\bf Q}=(h,h,-1.8)$ within the $(hhl)$ zone.  The inset
shows the scan direction within the $(h,k,-1.8)$ zone.}
\label{fg:ybco} \end{figure} 

For the calculated curve, the filling in of weight between the IC peaks occurs
because $\xi$ is less than the stripe spacing $na$.  In this case the weak
correlation between neighboring antiphase domains results in little
interference.  At ${\bf Q}_{\rm AF}$ the scattering from an individual domain
does not get cancelled by the weak contributions from neighboring domains.  In
contrast, the situation for La$_{1.85}$Sr$_{0.15}$CuO$_4$ is $\xi>na$.  That is
the necessary for condition for obtaining well resolved IC peaks.  Although the
correlations between antiphase domains are perhaps less well defined in 123, the
smaller value of $\xi$ suggests greater stripe fluctuations, a condition that
appears to correlate with an increased $T_c$ \cite{tran97a}.

\section{DISCUSSION}

It has been shown that the ${\bf Q}$ dependence of scattered intensity from
low-energy spin fluctuations in La$_{1.85}$Sr$_{0.15}$CuO$_4$ and
YBa$_2$Cu$_3$O$_{6.6}$ can be consistently interpreted on the basis of dynamic
antiphase antiferromagnetic domains.  The correlation lengths $\xi$ used to
model the scattering measurements were calculated from experimental values of
$\Gamma$ and $\hbar c_0$.  It was pointed out that the IC peaks become
unresolvable when $\xi<na$, where $na$ is the charge-stripe spacing.

There are, of course, other neutron scattering results that are consistent with
the stripe picture.  A comparison of $\tilde{\chi}''(\omega)$ in
superconducting YBa$_2$Cu$_3$O$_{6.6}$ and antiferromagnetic
YBa$_2$Cu$_3$O$_{6.15}$ shows that, despite a difference in $\omega$
dependence, the overall weight within the range 5 to 50~meV is comparable
\cite{sham93}.  Similar results were found in a comparison of
La$_{1.85}$Sr$_{0.15}$CuO$_4$ and La$_2$CuO$_4$ \cite{hayd96a}, although the
spectral weight for the superconductor is somewhat reduced when the comparison
is extended up to 300~meV.  With a strong segregation of holes of density
$n_h$, the density of Cu moments contributing to the low-energy ($<300$~meV)
spin fluctuations is roughly $1-2n_h$, so one may expect to see relatively
little change in overall spectral weight compared to the undoped antiferromagnet.

Measurements on Zn-doped La$_{1.86}$Sr$_{0.14}$CuO$_4$
have shown that the positions and $Q$-widths of the IC peaks are essentially
unaffected by the presence of the Zn \cite{mats93}.  Similarly, doping Zn into
YBa$_2$Cu$_3$O$_{6.6}$ does not change the ${\bf Q}$ dependence of the
scattering, but it does shift spectral weight down to low-energy, so that
$\tilde{\chi}''(\omega)$ looks more like that of an antiferromagnet
\cite{kaku93}.  This suggests that doping Zn into the CuO$_2$ planes 
destroys superconductivity by pinning the stripes.  

Recently there has been considerable excitement over observations of resonant
enhancements of the magnetic scattering below $T_c$ \cite{maso96,fong97}.  If
the enhanced scattering is assumed to come from intinerant electrons, then it
is certainly quite remarkable.  On the other hand, if the magnetic scattering
all comes from antiferromagnetic domains between charge stripes, the change in
the scattering might in some way be associated with a reduction in damping of
the spin fluctuations due to the onset of superconductivity.  In any case, it
will be interesting to see how models of these effects evolve.


\end{document}